\documentclass[12pt]{article}

\usepackage{amsmath, amssymb}

\input epsf.sty
\newcommand{\be}{\begin{equation}}
\newcommand{\ee}{\end{equation}}
\newcommand{\ben}{\begin{eqnarray}\displaystyle}
\newcommand{\een}{\end{eqnarray}}

\newcommand{\sectiono}[1]{\section{#1}\setcounter{equation}{0}}

\def\sqr#1#2{{\vcenter{\vbox{\hrule height.#2pt
       \hbox{\vrule width.#2pt height#1pt \kern#1pt
          \vrule width.#2pt}
       \hrule height.#2pt}}}}

\def\ifig#1#2#3{\xdef#1{fig.~\the\figno}
\writedef{#1\leftbracket fig.\noexpand~\the\figno}
\global\advance\figno by1}

\def\sqr#1#2{{\vcenter{\vbox{\hrule height.#2pt
       \hbox{\vrule width.#2pt height#1pt \kern#1pt
          \vrule width.#2pt}
       \hrule height.#2pt}}}}

\long\def\@makecaption#1#2{%
\vskip\abovecaptionskip
\sbox\@tempboxa{#1: #2}%
\ifdim \wd\@tempboxa >\hsize
  #1: #2\par
\else
  \global \@minipagefalse
  \hb@xt@\hsize{\box\@tempboxa\hfil}%
\fi
\vskip\belowcaptionskip}

\begin{document}

{}~ \hfill\vbox{
}\break

\vskip 1cm

\begin{center}

\large{\bf
A Phase Transition between Small and Large Field\\ Models of Inflation
}

\vspace{10mm}

\normalsize{Nissan Itzhaki and Ely D. Kovetz\\\vspace{2mm}{\em
Tel-Aviv University, Ramat-Aviv, 69978,
Israel}\\\vspace{2mm}nitzhaki@post.tau.ac.il,~elykovetz@gmail.com}

\end{center}\vspace{10mm}

\begin{abstract}

\medskip

We show that models of inflection point inflation exhibit a phase transition from a region in parameter space where they are of large field type to a region where they are of small field type. The phase transition is between a universal behavior, with respect to the initial condition, at the large field region and non-universal behavior at the small field region. The order parameter is the number of e-foldings. We  find integer critical exponents at  the transition between the two phases.

\end{abstract}

\newpage
\baselineskip=18pt

\sectiono{Introduction }
In this paper we present and study a phase transition (PT) phenomenon in models of cosmic inflation. The PT takes place when we interpolate in a specific way  between models of  large field inflation and small field inflation.

We shall use two definitions to make the  distinction between large and small field models and refer to them later as first and second categorizations:
\begin{enumerate}
\item[I.] Potentials in which the slow roll conditions for the canonically normalized inflaton field $\phi$ are satisfied over a distance in field space which is large compared to the four-dimensional Planck mass scale
  are called "large field" potentials, while potentials with only a small region where the slow roll conditions are met are "small field" potentials.
\item[II.] Adopting the categorization first introduced in \cite{DodKinKolb}, we distinguish between potentials according to their curvature,  $V''$,  at the slow roll region. Potentials satisfying $V''>0$ are called "large field" potentials, while "small field" models have $V''<0$ and "linear field" models satisfy $V''=0$.
\end{enumerate}
These two categorizations, though in some cases complementary, are not equivalent. For example, the recent model  \cite{Silverstein:2008sg} involving a $\phi^{\frac{2}{3}}$ potential is considered a large field model according to the first categorization, but a small field model according to the second. Similarly the even more recent model of \cite{McAllister:2008hb} is a linear field model according to the second categorization and a large field model according to the first.
Nonetheless we shall make use of both categorizations later on.

Models of large and small field inflation are quite different:
\begin{itemize}
\item For single field models of inflation, the tensor-to-scalar ratio $r$ and the spectral index $n_s$ are related by:
\begin{equation}
r=\frac{8}{3}(1-n_s)+\frac{16}{3}\frac{V''}{V}.
\end{equation}
This relation shows that positive curvature models, i.e.\ large field models with $V''>0$ are expected to generate larger values for $r$ and can be constrained by its measurement through gravity wave experiments, while negative curvature small field models predict small values for $r$. This implies that experimentally models with $V''>0$ are more severely constrained \cite{wmap5}.
\item In large field models, where the slow roll region is extended, the initial conditions of the inflaton generically do not play an important roll (the field will either start in slow roll conditions or slow down enough in the slow roll region to allow for enough e-foldings to accumulate and since inflation erases all memory of what happened before the last e-foldings, the exact initial conditions play no important role). In small field models, however, unless the inflaton starts out near the slow roll region of the potential, it will overshoot it without inflating the universe. Simply put, the Hubble friction does not have enough time to slow down the inflaton at the slow roll region.
\item Constructing large field models in string theory is quite  difficult (see, however, \cite{Silverstein:2008sg, McAllister:2008hb}). Small field models are generally easier to construct, largely since the slow roll conditions can be satisfied accidently in a small region.
\end{itemize}

Since the two types of models are distinctly different, one might ask if the transition between them is smooth or not. Namely, is there a PT between the two? To explore this possibility we have to study the physics along a trajectory which interpolates between small field and large field inflation. The problem is that the space of all possible models of single field inflation has an infinite dimension, and so there are infinitely many trajectories from the region of small field inflation to the region of large field inflation. Thus to proceed in a meaningful way we have to consider a slice of this space that truncates the number of dimensions. In the present paper we consider the simplest possibility of doing so via a one dimensional slice which is defined in the following
way. We rescale $\phi$ in the potential $V(\phi)$
\be\label{rs} \phi \to \lambda \phi ,\ee
while keeping the kinetic term canonically normalized.
From the definition of the slow-roll parameters
\begin{equation}
\epsilon=\frac{1}{2}\left(\frac{V'}{V}\right)^2,~~~~~\eta=\frac{V''}{V},
\end{equation}
and our first categorization above, we see that for any non-singular potential, $\lambda\to\infty$ gives a small field model while $\lambda\to 0$ yields a large field model.

The total number of e-foldings obtained at the slow roll region (the region where $\epsilon <1$) is denoted by $N$. As we vary $\lambda$ we expect to find  the two extreme values
\begin{equation}
\lim_{\lambda\rightarrow\infty}N(\lambda)=0,~~~~~~~~
\lim_{\lambda{\rightarrow}0}N(\lambda)=\infty.
\end{equation}
This suggests that it is natural to think of $N(\lambda)$ as an order parameter between the two phases of large and small field inflation.

The main point of the paper is that contrary to what might be expected, in some cases, in which $V(\phi)$ is regular and has an inflection point, there is a PT at some {\it finite} value of $\lambda$ where $N$ suddenly becomes infinite. This discontinuous transition happens when the parameter is varied continuously, reminiscent of phase transitions of the second kind in Landau-Ginzburg theory. Moreover, we find a scaling behavior near the critical point of the form:
\begin{equation}\label{scaling}
N(\lambda)\sim\frac{1}{\left(\lambda-\lambda_{c}\right)^{\gamma}}~,~~~~~~~~\lambda<\lambda_{c}~.
\end{equation}
As is often the case with scaling behavior, (\ref{scaling}) is universal. The sense in which it is universal here is that it does not depend on the initial condition of the inflaton. Namely, near $\lambda_{c}$ any non-singular initial condition yields (\ref{scaling}). This is due to the attractor property of inflation.\footnote{ For a general discussion on the attractor mechanism see \cite{Liddle:2000cg}. The efficiency of the attractor mechanism in inflection point inflation  models was discussed recently in \cite{Rouzbeh,Rajeev}.} This scaling obeys an integer exponent scaling law as we later show and explain in detail.

The  paper is organized as follows. In section 2 we demonstrate a PT in models of inflection point inflation (IPI). Such models  have recently become popular in the context of inflation in string theory \cite{Baumann:2007np,Baumann:2007ah,Krause:2007jk,Panda:2007ie,us,Hertzberg:2007wc,Linde:2007jn,
Baumann:2008kq} and the MSSM  \cite{Allahverdi:2006,Bueno Sanchez:2006xk,Allahverdi:2007}. We show numerically that IPI models satisfy the scaling behavior described in \eqref{scaling} with $\gamma=1$.
In Section 3 we discuss generalizations of the PT phenomenon. We show that it extends to models with an approximate inflection point and check the influence of higher powers of $\phi^n$ in the potential. For example, it turns out that the critical exponent $\gamma$ of the scaling behavior directly depends on the lowest degree of non-vanishing derivative in the extremal point of the potential around which inflation occurs.
Section 4 includes an analytic treatment of the PT properties described in Sections 2 and 3.
Finally, in Section 5 we illustrate that a somewhat different scaling behavior also takes place in a scenario with a time-independent IPI potential supplemented by a time-dependent contribution which helps solve the initial condition problem \cite{us}. This model has a similar PT (also with integer scaling behavior near the critical PT point) that occurs at a critical value of the initial condition of the time-dependent contribution to the potential. We conclude in section 6.

\sectiono{Phase Transition in Inflection Point Inflation}

In this section we consider the basic example of a PT between models of small and large field inflation.
The example is of inflection point inflation (IPI) where the inflaton's potential  can be approximated near the inflection point, at $\phi=0$, by
\be\label{ApproxIPI} V(\phi)= V_{0}(1+\beta \phi^3) . \ee
The fact that these kind of models are special follows  from the second categorization above. According to this categorization, for any $\beta$ the models are of large field for sign$(\beta) \phi>0$  and small field for sign$(\beta)\phi<0$.
Indeed, as we shall see, what determines the transition is whether the inflaton spends most of the time at positive or negative values.

From the point of view of our first categorization above, what determines if the potential is of large or small field form is the value of $\beta$, which controls the shallowness of the potential near the inflection point. For $\beta<<1$, the potential is of large field form (it is very flat near the inflection point and so the slow roll conditions are satisfied over a large range in $\phi$ space) while for $\beta>>1$ it has small field form.

We wish to demonstrate the PT property numerically, by finding a finite $\lambda_c$ for which $N(\lambda_c)=\infty$.
From \eqref{ApproxIPI}, we see that taking  $\phi\to \lambda \phi$ and keeping the kinetic term fixed is equivalent to keeping $\phi$ fixed and transforming
\begin{equation}\label{beta_trans}
\beta\longrightarrow\lambda^3\beta.
\end{equation}
Therefore, to demonstrate a PT we need to find a finite  $\beta_{c}$.

The equations of motion are
\begin{eqnarray}\label{inflation_equations}
3H^2&=&\frac{1}{2}\dot{\phi}^2+V_{0}(1+\beta \phi^3),\cr
& & \cr
\ddot{\phi}&=&-3H\dot{\phi}-3 V_0 \beta \phi^2,
\end{eqnarray}
where as usual $H=\dot{a}/a$. The classical equations of motion are invariant under\footnote{This symmetry is broken at  the quantum mechanical level by the Planck scale.}
\be\label{tr} V_0\to \alpha^2 V_0,~~~~t\to t/\alpha.\ee
This transformation does not change the region (in field space) of slow roll and it leaves our order parameter, $N=\int Hdt$, invariant. Thus  it can be used to  fix $V_0=1$.

We are left with the task of solving (\ref{inflation_equations}), with $V_0=1$,
numerically and calculating $N$ (while the inflaton is in the slow-roll region) as a function of $\beta$ for different values of the initial condition. These results are summarized in Table 1 which shows that there is a critical point around $\beta_{c}\sim0.7744$ below which $N$ goes to infinity independently of the initial condition.

To make contact with critical behavior and second order phase transitions it is more convenient to work with $\frac{1}{N}$ (rather than $N$) as the order parameter. The reason is that  it changes continuously from zero to finite values, as shown in figure 1. Indeed this figure looks very familiar from discussions on second order phase transitions. We can go even further in improving the definition of the order parameter. As we approach the critical point, most of the e-foldings are obtained near the inflection point. Since in this limit $H$ is roughly constant near the inflection point what determines the behavior of $\frac{1}{N}$ is the amount of time the inflaton spends near the inflection point. This in turn is determined by the inflaton's velocity
at the inflection point
\begin{equation}
\frac{1}{N}\sim\dot{\phi}(\phi=0)\equiv\dot{\phi}_0.
\end{equation}
Therefore, it makes sense to work with $\dot{\phi}_0$ as the order parameter.
Moreover, since $\dot{\phi}_0$ is a local operator (while $1/N$ is a non-local operator)
it is easier to  numerically calculate $\dot{\phi}_0$.
In the remainder of the paper we shall use these alternatives as the order parameters for our analysis.
\begin{table}\label{Table_beta}
\begin{center}
\begin{tabular}{|c|c|c|c|c|c|}
\hline
$\beta\Big\backslash\phi_{init}$& $0.5$& $1$& $5$& $10$& $100$\\
\hline
$0.1$& $ \infty $& $ \infty $& $ \infty $& $ \infty $& $ \infty $\\
\hline
$0.7$& $ \infty $& $ \infty $& $ \infty $& $ \infty $& $ \infty $\\
\hline
$0.7744$& $ \infty $& $ \infty $& $ \infty $& $ \infty $& $ \infty $\\
\hline
$0.7745$& $ \infty $& $ \infty $& $38404.9097$& $38218.5366$& $38218.5365$\\
\hline
$0.8$& $ \infty $& $ \infty $& $48.3532$& $48.3529$& $48.3529$\\
\hline
$1$& $ \infty $& $ \infty $& $5.2579$& $5.2579$& $5.2579$\\
\hline
$5$& $0.96866$& $0.46226$& $0.35061$& $0.35061$& $0.35061$\\
\hline
\end{tabular}
\caption{The amount of e-foldings generated for different initial conditions (assuming $\dot{\phi}_{init}=0$) for different values of the free parameter $\beta$ (holding $V_0=1$). We see that below the critical value $\beta_{c}\sim0.7744$ inflation is infinite for all initial conditions.
}
\end{center}
\end{table}

\begin{figure}[t!]
\begin{picture}(220,250)(0,0)
\vspace{0mm} \hspace{0mm} \mbox{\epsfxsize=150mm
\epsfbox{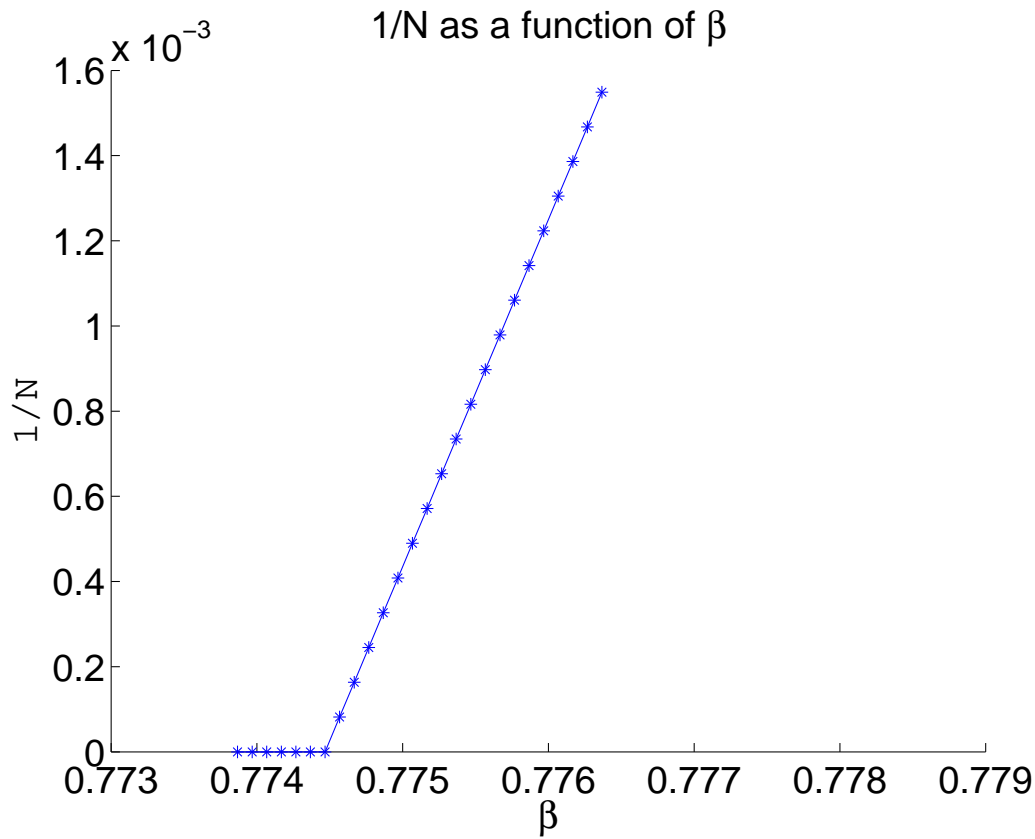}}
\end{picture}
\caption{Alternative for the order parameter of our PT. The behavior of $\frac{1}{N}$ as a function of $\beta$ is shown. The transition at the critical point is continuous, dropping to zero when it is crossed and inflation becomes infinite.}
\end{figure}

In Figure 2 we present the scaling behavior near $\beta_c$ both for $1/N$ and for $\dot{\phi}_0$.
We see that both cases exhibit scaling behavior when $\beta_c$ is approached with excellent accuracy. In addition, we see that the slope of all graphs shown is extremely close to $1$, which implies the relation
\begin{equation}\label{critexp}
\dot{\phi}_0\sim\frac{1}{N}\propto(\beta-\beta_{crit})^{1}.
\end{equation}
The fact that the critical exponent is an integer strongly suggests an analytic explanation. Such an explanation will be discussed in section 4. Before we turn to this explanation we consider in the next section some generalizations of this basic example.
\begin{figure}
\begin{picture}(220,250)(0,0)
\vspace{0mm} \hspace{0mm} \mbox{\epsfxsize=160mm
\epsfbox{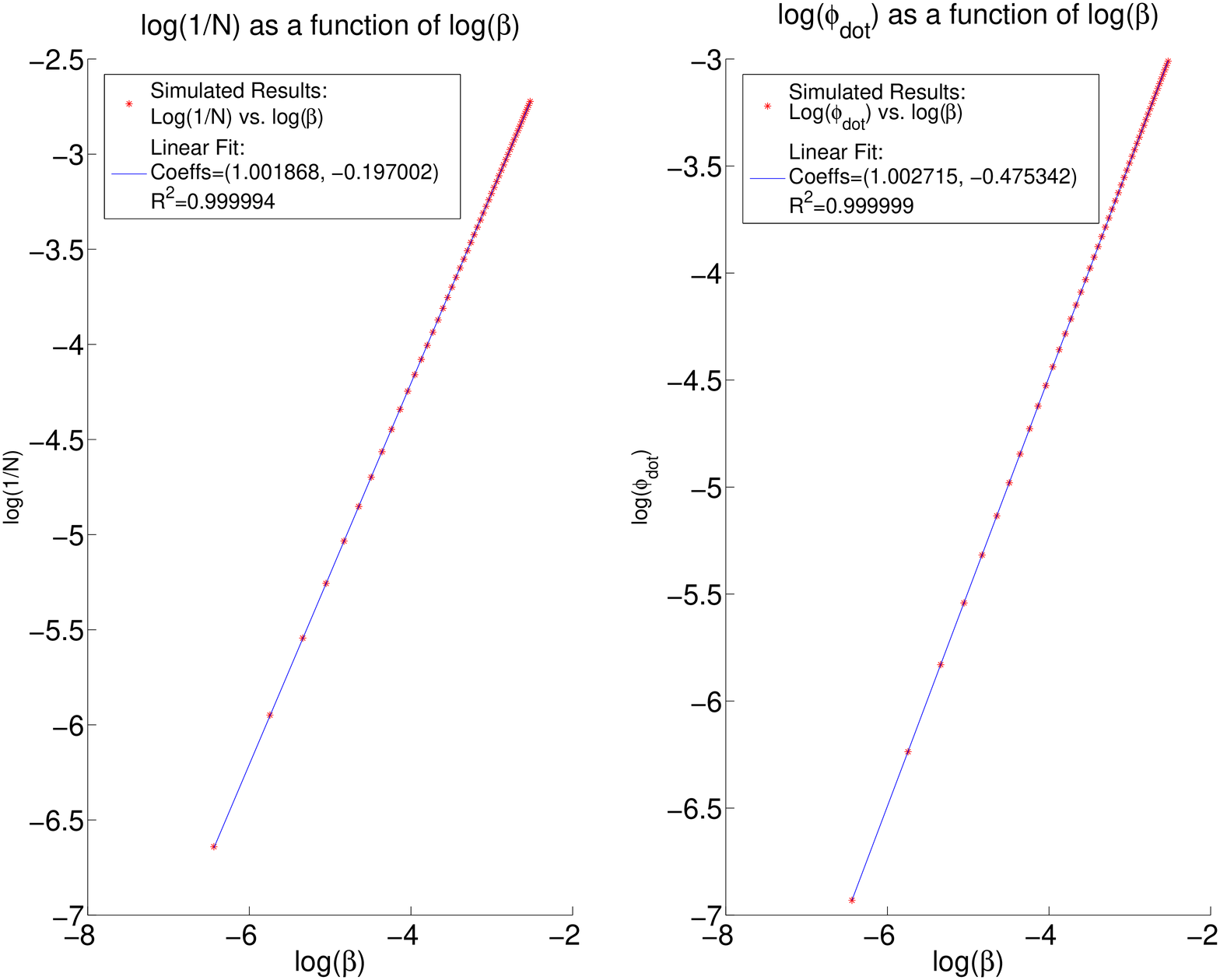}}
\end{picture}
\caption{Scaling of $\frac{1}{N}$ and $\dot{\phi}_0$ with distance of $\beta$ from its critical value. Both cases obey a scaling law with critical exponent $1$.}
\end{figure}

\sectiono{Irrelevant and Relevant Deformations}

In this section we consider some relevant and irrelevant deformations of the  potential (\ref{ApproxIPI}).

We start with the irrelevant deformations.
Consider the following potential.
\begin{equation}\label{trinomial}
V(\phi)=V_0\left(1+\beta\phi^3+\sum_{n_i>3}\alpha_i\phi^{n_i}\right).
\end{equation}
As before, we would like to study the PT as we rescale $\phi$ according to eq. (\ref{rs}). This rescaling is equivalent to
taking
\be \beta\to\lambda^3\beta,~~~~\alpha_i\to \lambda^{n_i}\alpha_i.\ee
Our numerical simulations show that, as expected,
the higher order terms are indeed irrelevant in the sense that they do not influence  the existence of the critical point and the  value of the critical exponent.

Only when the "marginal term" vanishes ($\beta=0$) the critical exponent is not $1$.
When the leading term in the potential is
\be\label{op} (1+\alpha \phi^{2n+1}),~~~n=1,2,...\ee
the critical exponent is $2n-1$.
Potentials in which the lowest non-zero derivative is even, i.e. of order $2n$, exhibit no critical behavior at all. We shall return to this in the analytic discussion in section 4.

\begin{figure}[t!]
\begin{picture}(220,250)(0,0)
\vspace{0mm} \hspace{0mm} \mbox{\epsfxsize=160mm
\epsfbox{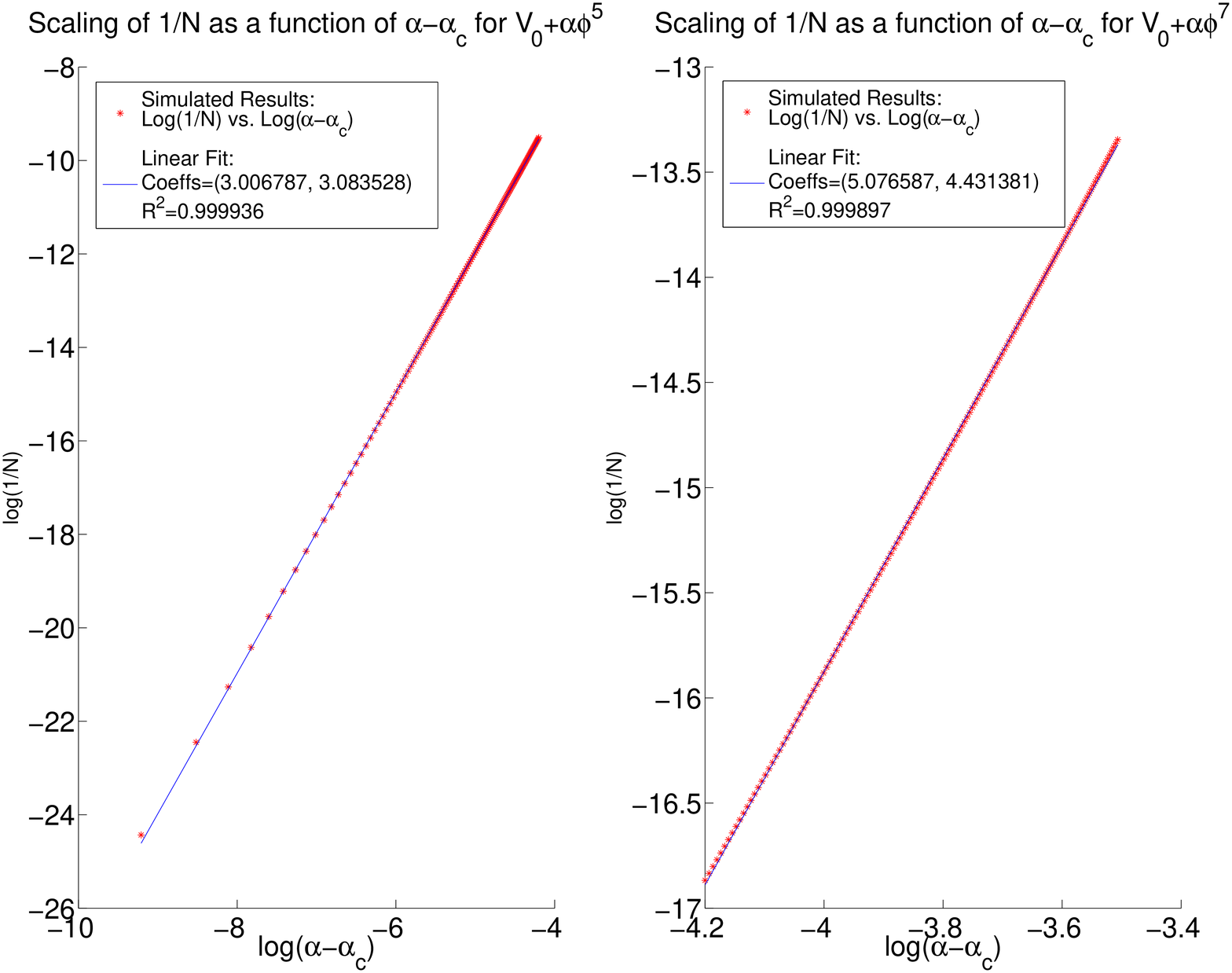}}
\end{picture}
\caption{Scaling of $\frac{1}{N}$ with distance of parameter from its critical value. {\it Left:} Approximated IPI potential with non-zero fifth derivative, critical exponent is 3. {\it Right:} Approximated IPI potential with non-zero seventh derivative, critical exponent is 5. }
\end{figure}

Next we consider a relevant deformation of the form
\begin{equation}
V=V_0(1+\delta\phi+\beta\phi^3).
\end{equation}
This time the rescaling (\ref{rs}) is equivalent to
\be\label{ro} \beta\to\lambda^3\beta,~~~~\delta\to \lambda\delta.\ee
It is easy to see that for any $\delta\neq 0$ there is no PT of the kind discussed before. Namely, there is no finite $\lambda_c$ for which $N(\lambda_c)=\infty$, and hence no scaling behavior of the form (\ref{scaling}). From the point of view of $\dot{\phi}_0$ as the order parameter this is clear since the minimal value of $\dot{\phi}_0$ is obtained in the slow roll approximation and is linear in $\delta$.
Therefore, we are never in the phase where the order parameter vanishes.

This can also be seen when considering $1/N$ as the order parameter.
The maximum number of e-foldings, obtained by tuning the initial condition, is \cite{Baumann:2007ah}
\be\label{max} N_{max}=\pi \sqrt{\frac{1}{3\delta \beta}}.\ee
Again we see that for any  finite value of $\delta$ the order parameter, $1/N$, never vanishes.

Still, as  is often the case in Landau-Ginzburg theory, some aspects of the PT remain. Here we find that for any $\delta\neq 0$ there is a $\lambda_c$ above  which $N$ depends on the initial condition, and to maximize $N$ we have to fine tune the initial condition. On the other hand for $\lambda<\lambda_c$ any non-singular initial condition gives $N=N_{max}$.
From (\ref{ro}) and (\ref{max}) we find the universal (with respect to the initial condition) behavior for $\lambda<\lambda_c$:
\be\label{hh} N(\lambda)=N(\lambda_c)\left(\frac{\lambda_c}{\lambda}\right)^2.\ee
The fact that $N(0)=\infty$ is trivial. What is interesting here is the fact that  (\ref{hh})  does not depend on the initial condition, and that it scales like $1/\lambda^2.$

\sectiono{Analytic Discussion}

In this section, our aim is to find an analytic explanation for the integer critical exponents we have found numerically. In order to do that, we have to simplify the equation of motion keeping only the most relevant features. Since near the critical point the inflaton spends most of the time near the inflection point, This means that as $\beta$ approaches $\beta_c$ we can approximate $H$ by a constant. This assumption allows us to replace our order parameter $\frac{1}{N}$ with $\frac{1}{t}$ where $t$ is the time spent by the inflaton in the vicinity of the inflection point (since $N_{e-folds}=\int\limits_{t_{start}}^{t_{end}}Hdt=H(t_{end}-t_{start})$).
We have also seen that irrelevant terms ( $\phi^n$ terms with $n>3$) are indeed irrelevant, in the sense that adding them to the potential does not modify the critical exponents. Therefore we expect to find a similar PT with the same critical exponents
for the following equation
\be \ddot{\phi}=-3H_0\dot{\phi}-3 V_0 \beta \phi^2,~~~ \mbox{with}~~~ 3H_0^2=V_0.\ee
This equation approximates (\ref{inflation_equations}) near the inflection point and is invariant under the same transformation  (\ref{tr}). Again we use this symmetry to fix $V_0$ (or $H_0$). This time we find it useful to fix $H_0=1/3$ which leaves us with
\be\label{al} \ddot{\phi}=-\dot{\phi}- \beta \phi^2.\ee
Unlike the original equation, (\ref{inflation_equations}), this equation is also invariant under
\be\label{k} \phi\to C\phi,~~~~\beta\to\beta/C^2.\ee
This means that solutions to (\ref{al}) with the initial condition $\phi(t=0)=\phi_0$ and $\dot{\phi}(t=0)=0$
are parameterized by the combination $\phi_0^2 \beta$ which is invariant under (\ref{k}). Thus to describe the PT we can either fix $\phi_0$ and vary $\beta$ or fix $\beta$ and vary $\phi_0$. We find it most convenient to fix $\beta=1$ and vary the initial condition $\phi_0$.

Our goal in the rest of this section is therefore to show that the equation
\be\label{nonlinosc} \ddot{\phi}=-\dot{\phi}-  \phi^2,\ee
exhibits a PT as a function of  $\phi_0$ and to find the relevant critical exponent.
To meet this goal we use the phase-space approach.\footnote{We are grateful to Barak Kol for suggesting this to us and for deriving some of the results below. This approach was used recently in \cite{Rajeev}.} We rename the axes as $x=\phi,~y=\dot{\phi}$ and  reduces the second order differential equation to a system of two first order ones
\begin{eqnarray}
\dot{x}&=&y,\cr
\dot{y}&=&-y-x^2.
\end{eqnarray}
By examining the derivative
\begin{equation}
\frac{dy}{dx}=\frac{\frac{dy}{dt}}{\frac{dx}{dt}}=\frac{-y-x^2}{y}
\end{equation}
we  deduce a few general conclusions:
\begin{itemize}
\item When crossing the $x$ axis ($y=0$) the slope is always infinite.
\item When crossing the $y$ axis ($x=0$) the slope is always $-1$.
\item When crossing the contour $y=-x^2$, the slope is always zero (except at $x=0$).
\item For negative $y$, outside the contour $y=-x^2$, the derivative is positive and so the trajectories are monotonically increasing.
\item For negative $y$, inside the contour $y=-x^2$, the derivative is negative and so the trajectories are monotonically decreasing.
\end{itemize}
This explains the trajectories in Figure 4, from which we see that indeed there is a critical initial condition point and that orbits starting at different points on the x-axis (i.e. initial conditions with zero velocity) either converge to the stable point at $x=y=0$ or overshoot it and asymptotically approach $x, y~\to-\infty$ .

These general considerations, however, are not sufficient to determine the critical exponent. To derive the critical exponent we have to
\begin{figure}
\begin{picture}(220,200)(0,0)
\vspace{0mm} \hspace{40mm} \mbox{\epsfxsize=80mm
\epsfbox{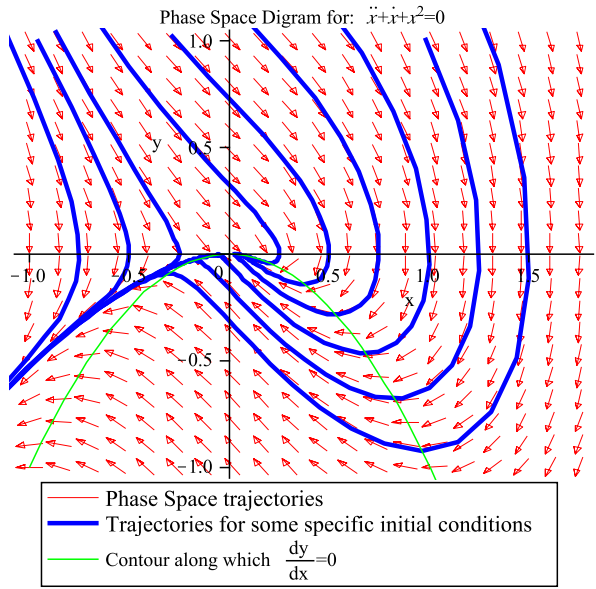}}
\end{picture}
\caption{Phase Space trajectories of solutions to \eqref{nonlinosc} for different initial conditions. }
\end{figure}
analyze the behavior near  $x=0$.
The region near $x=0$ can be approached in two different ways:

(I) We can neglect the $x^2$ term. Then the second order equation becomes linear, with the solutions
   \be x=c_1 e^{-t} +c_2.\ee
This corresponds in the original  problem to keeping the Hubble friction term while neglecting the slope of the potential.
Namely, we are in the de-Sitter approximation. This solution is valid to the right of $x=0$ and we see that for $c_2=0$ the trajectory will reach the origin $x=y=0$ at $t\rightarrow\infty$, coninciding with the {\it critical trajectory} beginning exactly at the critical point on the $x$ axis.

(II) We can also neglect the $\ddot{x}$ term.  Then the equation becomes first order and we have the solution
\be x=\frac{1}{t+c}.\ee
This corresponds in the original problem to taking the slow roll limit. This solution with $c=0$ describes a trajectory that starts at the origin at $t\to-\infty$ and moves to the left of $x=0$, coinciding with the curve along which the derivative $\frac{dy}{dx}$ vanishes (the curve $y=-x^2$ with $x<0$).


Now, in order to find the critical exponent, we need to consider the behavior of a trajectory that starts at some small distance $\epsilon'$ from the critical point on the $x$ axis. Such a trajectory will never reach the origin, but will eventually cross the $y$ axis below zero. However, when it approaches $x=0$, the critical trajectory is a good approximation for it (see Figure 5). Therefore, we start with solution (I)
\begin{equation}
x=ce^{-t}-\epsilon~~~,~~~y=-ce^{-t},
\end{equation}
parametrizing the distance from the $c2=0$ solution by $\epsilon$. Here it is important to note that since the equations are regular in this region, the relation between $\epsilon'$ and $\epsilon$ is linear. Therefore, finding a relation between $t$ near the $x=0$ and $\epsilon$ should give us the critical exponent.
The trajectory will cross the $x=0$ point at $y=-\epsilon$. We shall continue with this solution until it becomes unsatisfactory, at which point we will replace it with the second solution. The approximation (I) clearly breaks down when $\dot{x}\sim x^2\sim\epsilon^2$, which happens at
\begin{equation}
t_{*}\simeq\log\left(\frac{c}{\epsilon^2}\right)~~~,~~~x_{*}\simeq\epsilon^2-\epsilon\sim-\epsilon~~~,
~~~\dot{x}_{*}\simeq-\epsilon^2.
\end{equation}
Thus the first approximation is reliable up to $x\sim-\epsilon$, which, as is clear from Figure 5, is indeed in the vicinity of the curve along which approximation (II) is valid. However, as we saw above, the solution trajectory crosses this curve parallel to the $y$ axis and therefore must deviate from it (mush as it deviated from the critical trajectory to the right of $x=0$). We can therefore continue with the second approximation (I), starting at $x\sim-\epsilon$, until it, too, becomes invalid. To find when this occurs, we start with
\begin{equation}
x\sim-\epsilon-\delta x~~~,~~~\dot{x}\sim-\epsilon^2,
\end{equation}
and stop when $\ddot{x}\sim\dot{x}\sim\epsilon^2$, at which stage neglecting the $\ddot{x}$ term is no longer a good approximation. We find
\begin{equation}
\ddot{x}=\dot{y}=\frac{dy}{dx}\frac{dx}{dt}=\frac{-y-x^2}{y}y=-y-x^2=\not{\epsilon^2}-\not{\epsilon^2}-2\delta x\epsilon-(\delta x)^2\sim\delta x\epsilon
\end{equation}
which is no longer negligible once $\delta x\sim\epsilon$ or $x\sim-2\epsilon$. 
\\
We conclude that the time spent by the inflaton in the vicinity of the inflection point can be divided into regions with different characteristics: in the interval to the right of zero it behaves as $t_{right}\sim -\ln\epsilon$, while in the interval $[-2\epsilon,-\epsilon]$ it satisfies
\be\label{p} t_{left}\sim\frac{1}{\epsilon}.\ee
Since for $\epsilon<<1$ we have $|\ln\epsilon|<<\frac{1}{\epsilon}$, this implies that indeed the critical exponent in the relation between $\frac{1}{t}$ and $(x_{init}-x_{crit})$ is $\gamma=1$.

\begin{figure}
\begin{picture}(240,240)(0,0)
\vspace{0mm} \hspace{0mm} \mbox{\epsfxsize=160mm
\epsfbox{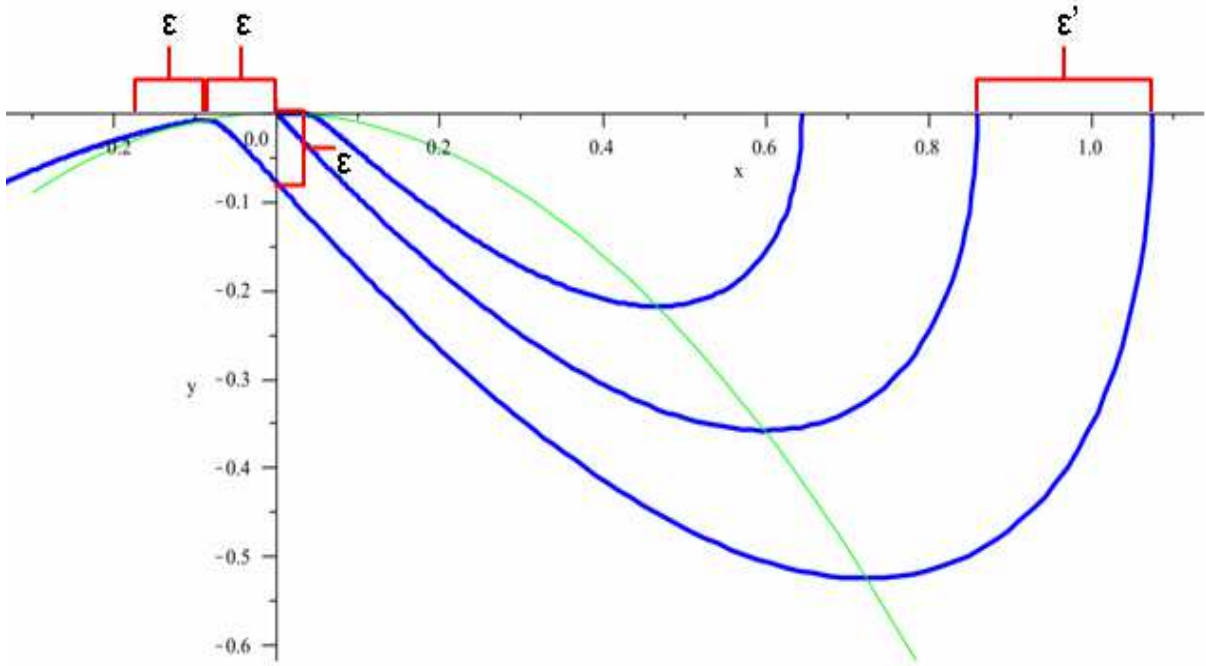}}
\end{picture}
\caption{Phase Space trajectories of solutions to \eqref{nonlinosc} for different initial conditions. The curve $y=-x^2$, along which the derivative $\frac{dy}{dx}=0$ is also plotted.}
\end{figure}

We end the section by considering very briefly the higher order cases of section 3. Using the same considerations as before we find that a potential of the form
(\ref{op}) leads to
\begin{equation}\label{x_4}
\ddot{x}+\dot{x}+x^{2n}=0.
\end{equation}
Drawing it in phase space (see Figure 6) we see the same qualitative behavior as for the previous potential. Repeating the approximations we did in the last section (solution (I) will obviously remain the same, while in solution (II) we will have $x\sim\frac{1}{t^2n-1}$) the logarithmic behavior to the right will remain, but to the left we will get the relation $x\sim\frac{1}{t^{2n-1}}$ which indeed explains a scaling law with an integer exponent of $2n-1$.
\begin{figure}
\begin{picture}(200,220)(0,0)
\vspace{0mm} \hspace{0mm} \mbox{\epsfxsize=80mm
\epsfbox{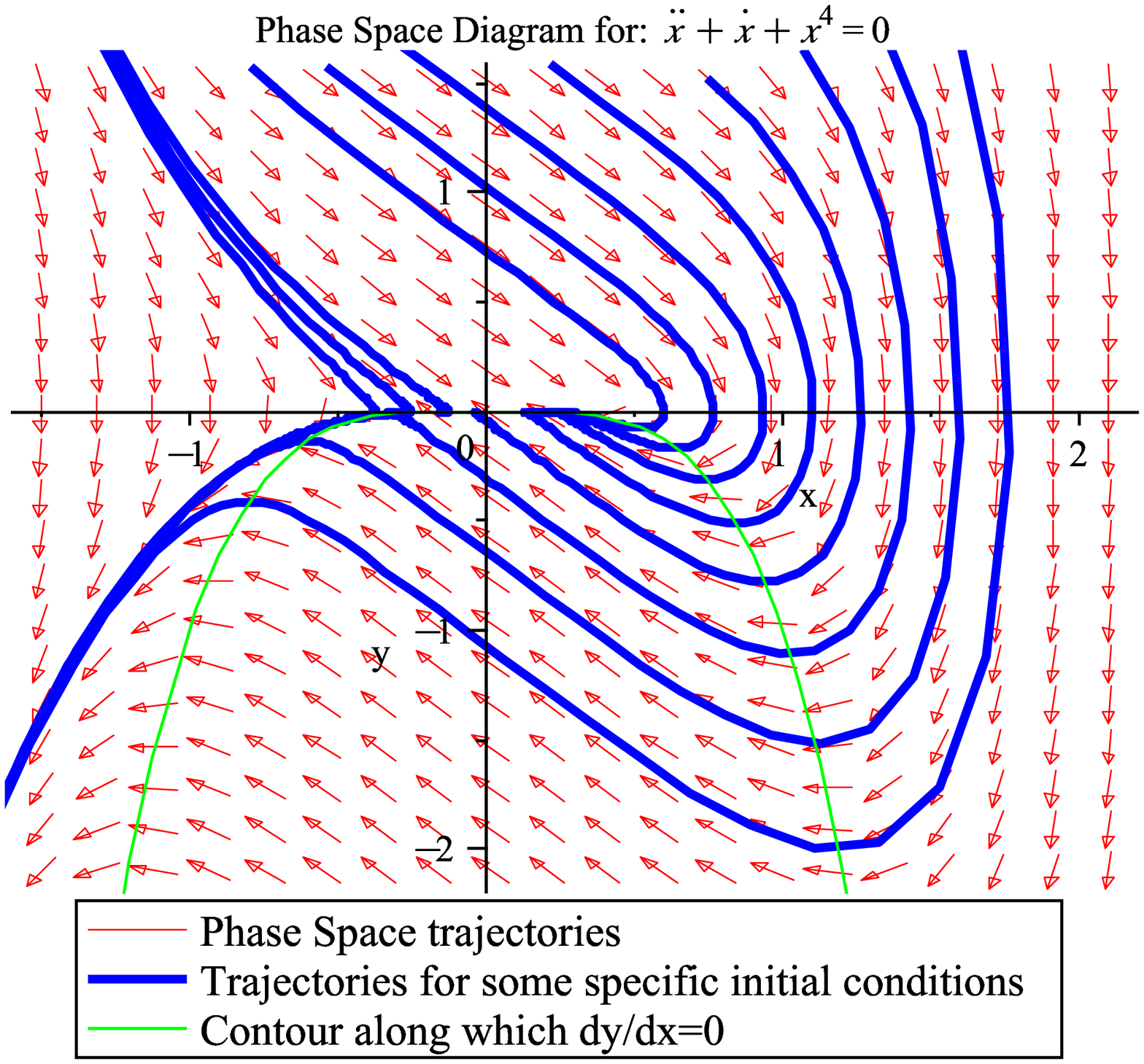}}
\end{picture}
\begin{picture}(200,220)(0,0)
\vspace{0mm} \hspace{0mm} \mbox{\epsfxsize=75mm
\epsfbox{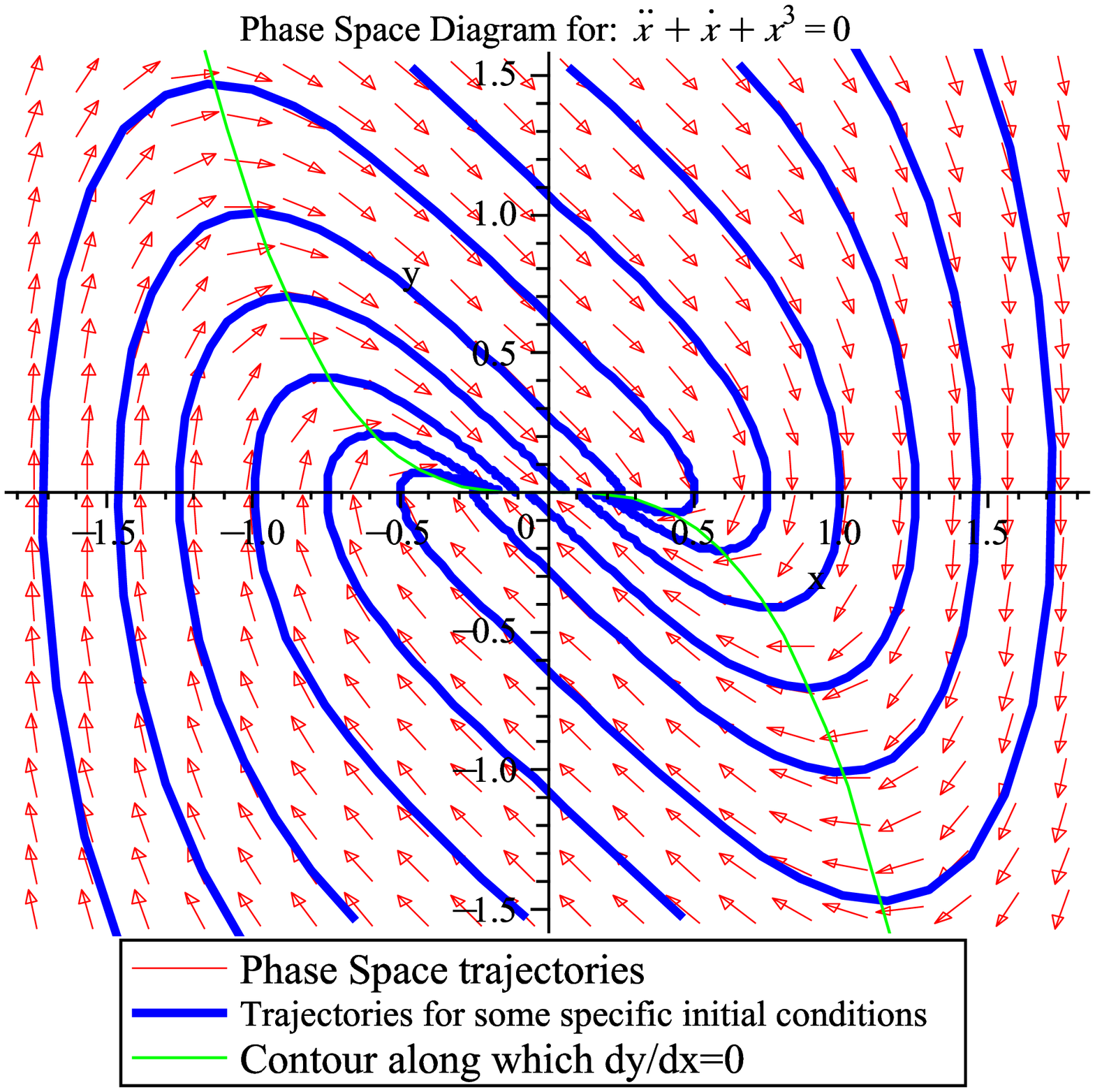}}
\end{picture}
\caption{Phase Space trajectories of solutions to \eqref{x_3} and \eqref{x_4} for different initial conditions.}
\end{figure}

On the other hand, for potentials with $\phi^{2n}$ as the lowest non-zero terms we get
\begin{equation}\label{x_3}
\ddot{x}+\dot{x}+x^{2n-1}=0.
\end{equation}
Note that this equation, unlike (\ref{x_4}),  is invariant under $x\to -x$. This implies that there is no
PT. In Figure 7 this is illustrated for $n=3$ where we can see  that no initial condition leads to a critical solution.

\sectiono{PT in a Time-Dependent IPI Model}

So far we have discussed the appearance of phase transitions in models of time-independent IPI. In this section we show that this phenomenon extends to cases of time-dependent potentials, although in this case the critical parameter is not a parameter in the potential or the initial condition for the inflaton, as we shall momentarily see. To demonstrate this, we focus on a stringy inspired time-dependent setup that was introduced in \cite{us}. The relevant equations of motion are
\ben\label{itr} 3H^2&=&\frac12 \dot{\phi}^2 + V_{static} +V_{dynamic}, \nonumber\\
\dot n&=&-3 H n,\\
\ddot{\phi}+3 H \dot{\phi}& =&
-\frac{d}{d\phi}\left( V_{static} +V_{dynamic}\right),\nonumber
\een
where $V_{static}$ is a small field potential (according to the first categorization) with an inflection point.

As is clear from the discussion above, the problem with such a $V_{static}$ is that for a generic initial condition the inflaton will overshoot the inflection point and the universe will not inflate. The resolution proposed in \cite{us} to the problem\footnote{Similar resolutions were proposed in the context of the overshoot problem associated with moduli stabilization \cite{overshoot} in \cite{krs}.} is that particles with mass that depends on the expectation value of the inflaton, and satisfies
\be\label{tt} \frac{d V_{static}}{d\phi} \frac{d M}{d\phi}< 0,\ee
induce a time-dependent potential
\be V_{dynamic}=n(t) M(\phi)\ee
that slows down the inflaton at the inflection point and resolves the overshoot problem (see Figure 7 for more details).

In \cite{us} it was observed that the initial value of $n_0$ needed in order to generate a large number of e-foldings is quite small. Here we make the point that there is a critical value of $n_0$ which we denote by $n_c$, above which $N=\infty$. In addition, we find that as $n_0$ approaches $n_c$ from below we obtain a scaling behavior
\be N\sim \frac{1}{n_c -n_0},\ee
which implies that in this case the critical exponent $1$ is also obtained with respect to the initial condition. 

Before discussing the numerical results, let us present some of the technical details.
The potentials we worked with are the ones of \cite{us} that are typical in stringy compactifications.
\begin{eqnarray}\label{static}
V_{static}&=&a_1 \exp(- j_1\alpha \phi) +a_2\exp(- j_2\alpha \phi)  +a_3 \exp(-j_3\alpha\phi) \cr
& & \cr
j_1&=&12,~~ j_2~=~10,~~ j_3~=~8,~~~\alpha=\frac{1}{\sqrt{24}},~~ \cr
& &\cr
\frac{a_1}{a_3}&=&\frac23L_{inflection}^4,~~\frac{a_2}{a_3}~=~ -\frac85L_{inflection}^2,~~,
\end{eqnarray}
where $L_{inflection}=e^{\frac{\phi_{inflection}}{\sqrt{24}}}$, and $L$ is the size of the compact manifold. The particles with masses that depend on the inflaton and satisfy (\ref{tt}) are branes that wrap the compact manifold. They induce
\be\label{par} V_{dynamic}=n(t)\exp(3\alpha\phi) .\ee
(We took the initial condition to be $\phi=\dot{\phi}=0$, which corresponds to $L=1, \dot{L}=0$.)

\begin{figure}
\begin{picture}(110,86)(0,0)
\vspace{0mm} \hspace{0mm} \mbox{\epsfxsize=37mm
\epsfbox{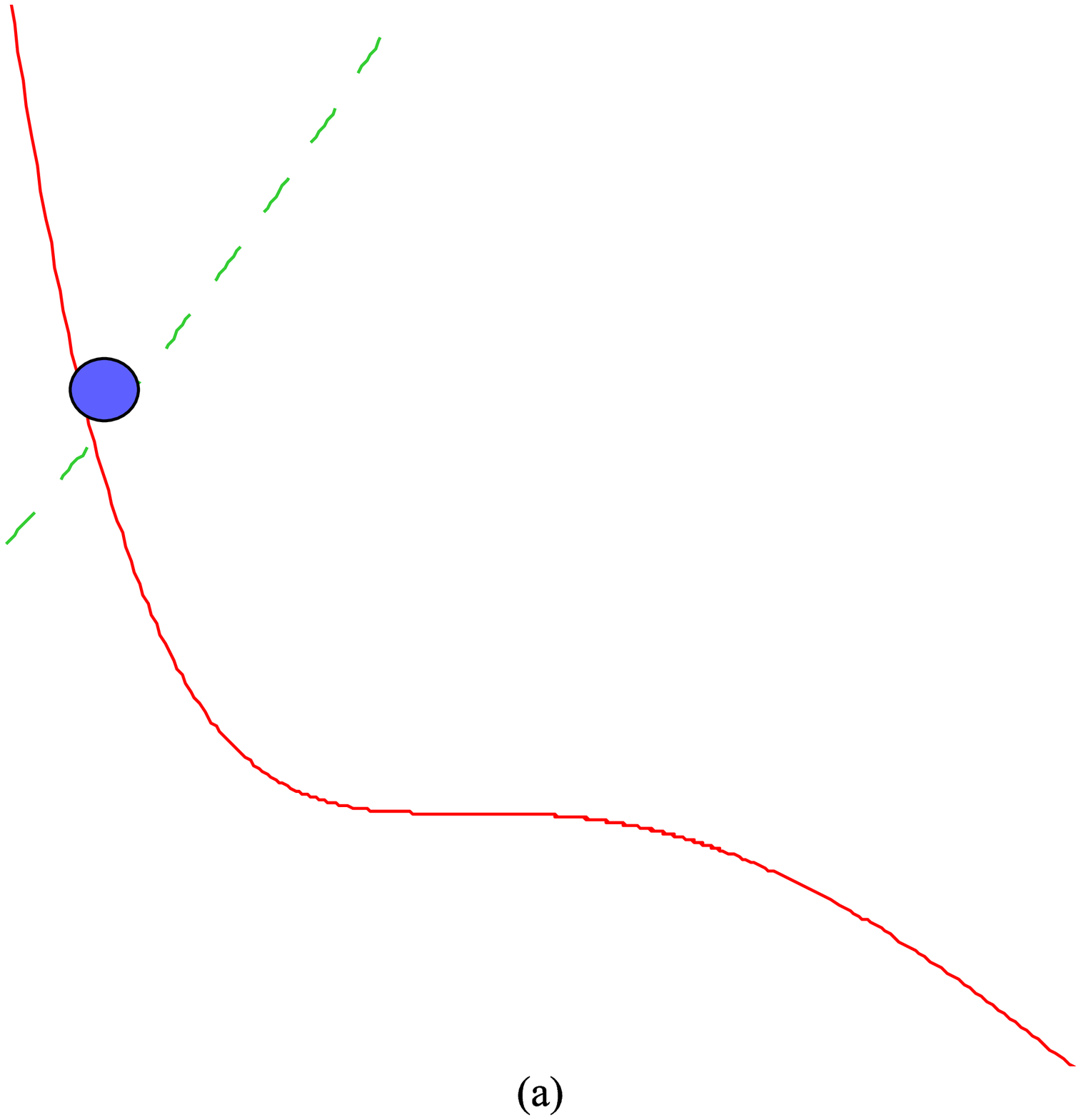}}
\end{picture}
\begin{picture}(110,86)(0,0)
\vspace{0mm} \hspace{0mm} \mbox{\epsfxsize=37mm
\epsfbox{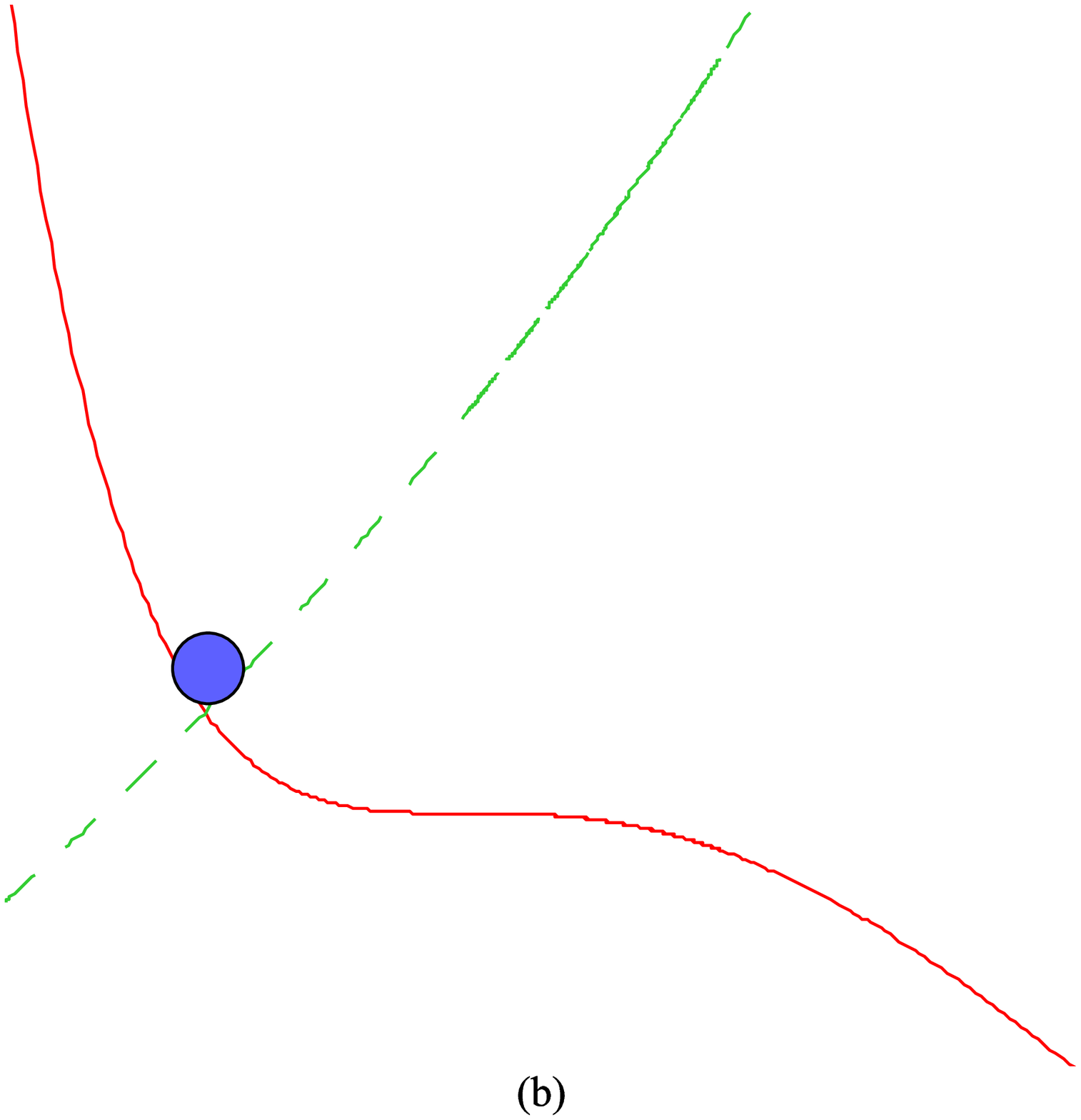}}
\end{picture}
\begin{picture}(110,86)(0,0)
\vspace{0mm} \hspace{0mm} \mbox{\epsfxsize=37mm\epsfbox{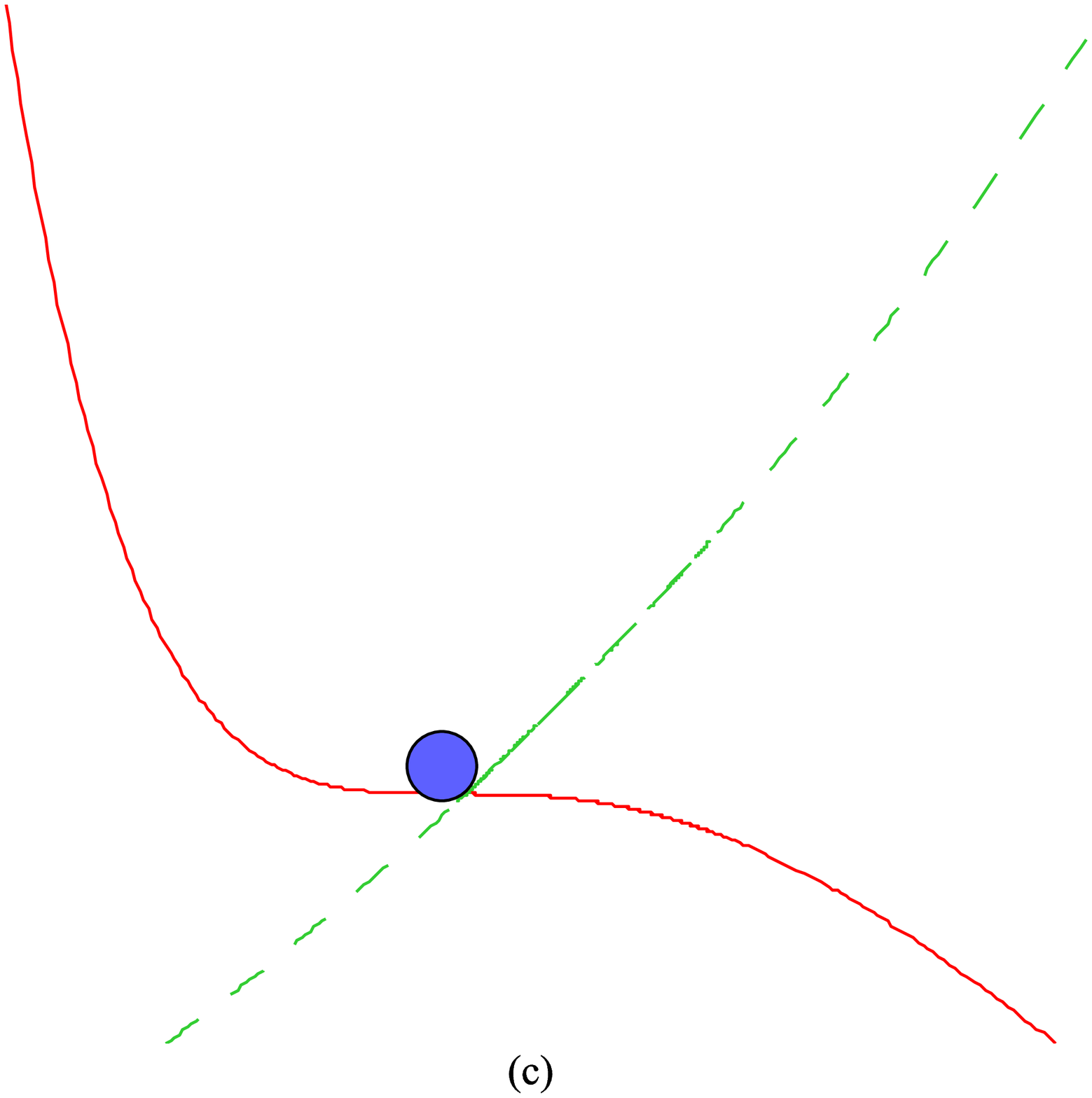}}
\end{picture}
\begin{picture}(110,86)(0,0)
\vspace{0mm} \hspace{0mm} \mbox{\epsfxsize=37mm
\epsfbox{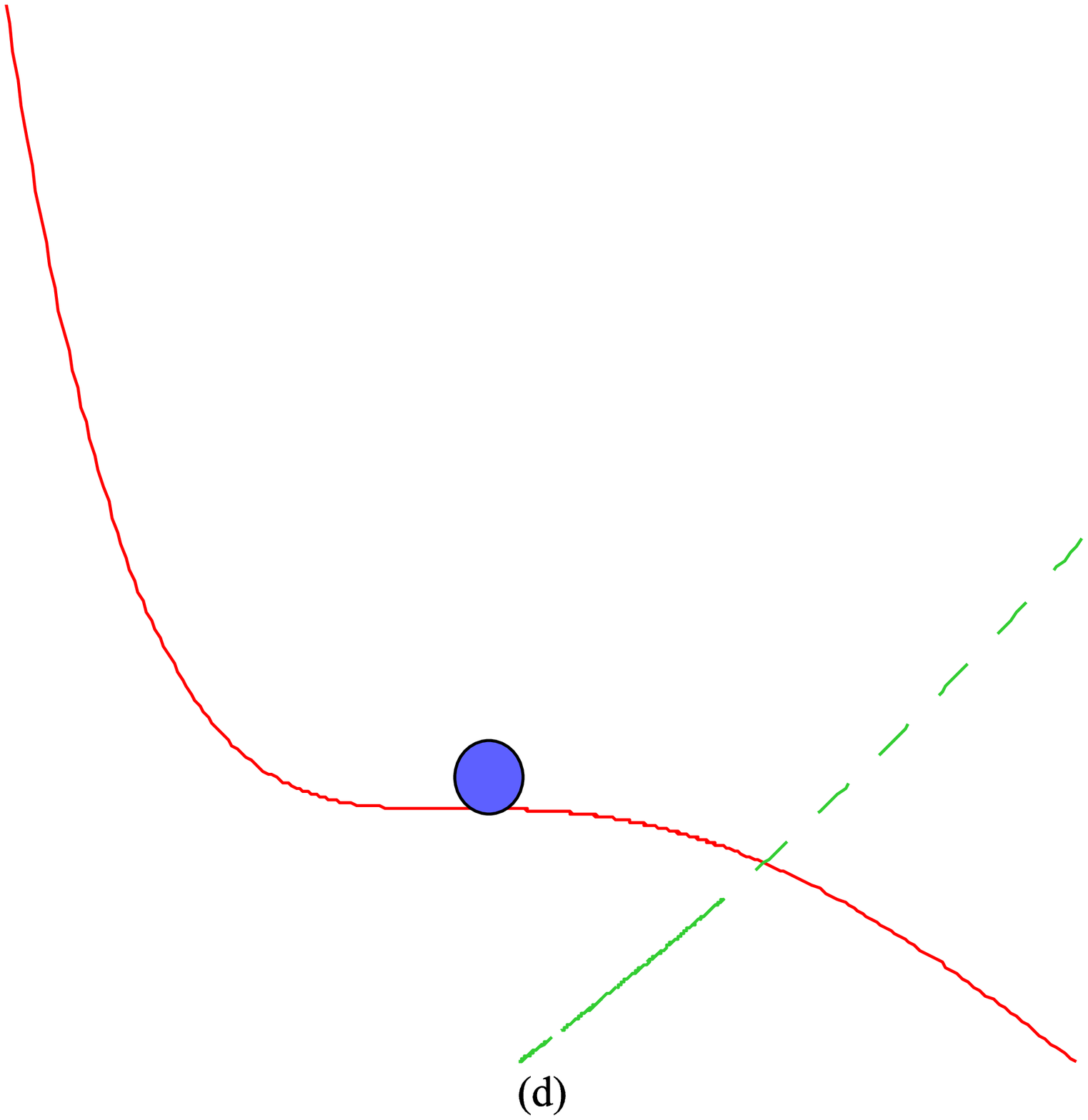}}
\end{picture}
\caption{ A heuristic demonstration of how a time dependent
potential, that scales like $1/a^3(t)$, can resolve the overshoot
problem of IPI. The static piece of the potential is denoted by the
red/solid line and the time dependent piece of the potential by the
green/dashed line. In (a),(b) the static potential is steep, but
since there has not been significant expansion the time-dependant
potential is able to balance it, preventing $\phi(t)$ from acquiring
a large velocity. Heuristically, $\phi(t)$ (denoted by the
blue/filled circle) follows the dilution of the time dependent
potential. In (c) $\phi(t)$ enters the shallow region (where the
slow roll parameters are small) where inflation takes place and the
time dependent potential starts to slow down exponentially. In (d)
$\phi(t)$ is dominated solely by the static potential but its
velocity is now low enough to allow prolonged inflation.}
\end{figure}

Now, solving the equations above numerically, we find a critical initial condition for $n(t)$, above which the inflaton is stuck at the inflection point (see table below).
\begin{table}
\begin{center}
\begin{tabular}{|c|c|c|c|c|c|}
\hline
$L_{inflection}\Big\backslash{n_{0}}$& $1.5e-05$& $0.000315$& $0.00081$& $0.00541$& $0.099035$\\
\hline
$2$& $0.46215$& $0.46474$& $0.46909$& $0.51401$& $51130.0048$\\
\hline
$5$& $0.42172$& $0.45187$& $0.51436$& $41.6388$& $\infty$\\
\hline
$8$& $0.42516$& $0.89218$& $44.8478$& $\infty$& $\infty$\\
\hline
$10$& $0.43725$& $53.3611$& $\infty$& $\infty$& $\infty$\\
\hline
$20$& $384.1207$& $\infty$& $\infty$& $\infty$& $\infty$\\
\hline
\end{tabular}
\caption{$N$ as a function of  $n_{0}$ and  $L_{inflection}$. We see that for any $L_{inflection}$ there is a critical value of $n_0$ above which $N=\infty$.}
\end{center}
\end{table}
Near the critical point, observing $t$ (the time spent near the inflection point) instead of $N$, as we did in Section 4, we notice the same scaling behavior, obeying a critical exponent of $1$. This is plotted in Figure 8. Our simulation shows that the critical exponent does not depend on the details of $V_{static}$ and $V_{dynamic}$. For example, replacing the $3$ in  (\ref{par}) by, say, $2$ would change $n_c$ but will not change the critical exponent.

\begin{figure}
\begin{picture}(220,220)(0,0)
\vspace{0mm} \hspace{35mm} \mbox{\epsfxsize=90mm
\epsfbox{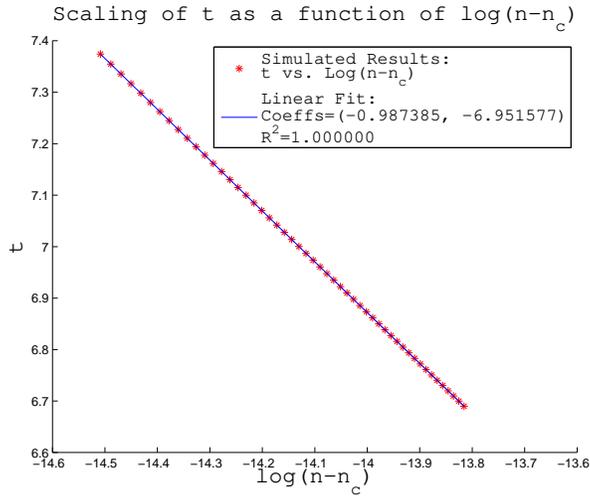}}
\end{picture}
\caption{Scaling behavior of the string Theory model near the critical point in the initial condition of the time-dependent contribution to the potential. Plotting the time spent in vicinity of inflection point vs. the Log of distance from critical point, we see the same scaling behavior encountered before, obeying a critical exponent close to $1$.}
\end{figure}

\sectiono{Conclusions}

The goal of the paper was to show that in IPI models, that have become quite popular in string theory recently,  the efficiency of the attractor mechanism varies in a discontinuous  fashion as we vary various parameters  continuously (as happens in phase transitions of the second king in Landau-Ginzburg theory). The parameter space of these models can be divided into two regions. In one (the small field region), $N$ depends on the initial condition and generally speaking, to obtain a large $N$ the initial condition should be fine tuned. In the large field region the efficiency of the attractor mechanism is maximized and all non-singular initial conditions yield the same $N$. The surprising aspects of our findings are:\\
(I) The transition between the two regions happens at {\it finite} values of the parameters that define the potential.\\
(II) In some cases critical behavior takes place at the transition between the two regions, obeying an integer exponent scaling law, which is shown and explained above.\\
These findings shed new light on the approach one should take in examining an IPI model in terms of the required fine tuning of the parameters of the model and the initial conditions. A good example is the time-dependent string theory model, where it is now evident that less care in fine tuning of the initial density of the stringy particles is needed in order for the proposed mechanism to work, since below a specific (and quite low) critical value, all values are equally sufficient. 

\hspace{5mm}

\noindent {\bf Acknowledgements}

We thank Ofer Aharony, Eyal Heifetz, Philip Rosenau and Barak Kol  for discussions. Barak Kol's crucial contribution to section 4 is much appreciated. This work is supported in part by the Israel Science Foundation (grant number 1362/08) and by the European Research Council  (grant number 203247).

\end{document}